\documentclass[twocolumn]{aastex63}

\received{May 29, 2020}
\revised{July 3, 2020}
\accepted{\today}
\submitjournal{ApJL}

\shorttitle{Fe\,{\sc i} emission in the day-side specturm of WASP-33b}
\shortauthors{S.K. Nugroho et al.}
\graphicspath{{./}{}}

\begin{document}

\title{Detection of Fe\,{\sc i} Emission in the Day-side Spectrum of WASP-33b\footnote{Based on data collected at Subaru Telescope, which is operated by the National Astronomical Observatory of Japan.}}

\correspondingauthor{Stevanus K. Nugroho}
\email{s.nugroho@qub.ac.uk, skristiantonugroho@gmail.com}

\author[0000-0003-4698-6285]{Stevanus K. Nugroho}
\affiliation{School of Mathematics and Physics, Queen's University Belfast, University Road, Belfast, BT7 1NN, United Kingdom}
\author[0000-0002-9308-2353]{Neale P. Gibson}
\affiliation{School of Physics, Trinity College Dublin, The University of Dublin, Dublin 2, Ireland}
\author[0000-0001-6391-9266]{Ernst J. W. de Mooij}
\affiliation{School of Mathematics and Physics, Queen's University Belfast, University Road, Belfast, BT7 1NN, United Kingdom}
\author[0000-0003-3636-5450]{Miranda K. Herman}
\affiliation{Astronomy $\&$ Astrophysics, University of Toronto, 50 St. George St., Toronto, ON M5S 3H4, Canada}
\author[0000-0002-9718-3266]{Chris A. Watson}
\affiliation{School of Mathematics and Physics, Queen's University Belfast, University Road, Belfast, BT7 1NN, United Kingdom}
\author[0000-0003-3309-9134]{Hajime Kawahara}
\affiliation{Department of Earth and Planetary Science, The University of Tokyo, Tokyo 113-0033, Japan}
\affiliation{Research Center for the Early Universe, School of Science, The University of Tokyo, Tokyo 113-0033, Japan}
\author{Stephanie Merritt}
\affiliation{School of Mathematics and Physics, Queen's University Belfast, University Road, Belfast, BT7 1NN, United Kingdom}

\begin{abstract}
We analyze the high-resolution emission spectrum of WASP-33b taken using the High Dispersion Spectrograph (R\,$\approx$\,165,000) on the 8.2-m Subaru telescope. The data cover $\lambda$\,$\approx$\,$6170$-$8817$\,\AA, divided over 30 spectral orders. The telluric and stellar lines are removed using a de-trending algorithm, {\sc SysRem}, before cross-correlating with planetary spectral templates. We calculate the templates assuming a 1-D plane-parallel hydrostatic atmosphere including continuum opacity of bound-free H$^{-}$ and Rayleigh scattering by H$_{2}$ with a range of constant abundances of Fe\,{\sc i}.
Using a likelihood-mapping analysis, we detect an Fe\,{\sc i} emission signature at 6.4-$\sigma$ located at $K_{\mathrm{p}}$ of 226.0\,$^{+2.1}_{-2.3}$\,km\,s$^{-1}$and $v_{\mathrm{sys}}$ of -3.2\,$^{+2.1}_{-1.8}$\,km\,s$^{-1}$ -- consistent with the planet's expected velocity in the literature. We also confirm the existence of a thermal inversion in the day-side of the planet which is very likely to be caused by the presence of Fe\,{\sc i} and previously-detected TiO in the atmosphere. This makes WASP-33b one of the prime targets to study the relative contributions of both species to the energy budget of an ultra-hot Jupiter.
\end{abstract}

\keywords{Exoplanet atmospheres (487); Exoplanet atmospheric composition (2021); High resolution spectroscopy (2096)}


\section{Introduction} \label{sec:intro}
Thermal inversions in hot Jupiters have been predicted by \citet{Hubeny2003} and \citet{Fortney2008} due to the presence of strong optical absorption of incoming stellar radiation, in particular from TiO/VO. There are now many hot Jupiters with a detected stratosphere \citep[e.g.][]{ Arcangeli2018, Mansfield2018, Haynes2015, Evans2017}, however up until now, TiO was only detected in the atmosphere of WASP-33b \citep{Nugroho2017} and WASP-19b (\citealt{Sedaghati2017}; although see also \citealt{Espinoza2019}). This raises the question of whether other species can cause temperature inversions.

One particular scenario suggests that for a planet with equilibrium temperature (T$_{\mathrm{eq}})>$ 2200\,K around an early-type star, thermal inversion layers could still exist even without TiO/VO, as other strong optical opacity sources can absorb sufficient incoming stellar radiation, such as Fe\,{\sc i} and continuum H$^{-}$ \citep{Lothringer2018, Lothringer2019}. The signature of Fe\,{\sc i} has since been detected in the transmission spectrum of several hot Jupiters, including KELT-9b \citep{Hoeijmakers2018}, WASP-121b \citep{gibson2020, Cabot2020, Bourrier2020}, WASP-76b \citep{Ehrenreich2020}, and KELT-20b/MASCARA-2b \citep{Nugroho2020, Stangret2020, Hoeijmakers2020}. Recently, \citet{Pino2020} detected Fe\,{\sc i} in the day-side spectrum of KELT-9b for the first time using the data taken with HARPS-N on the Telescopio Nazionale Galileo. They also showed that the Fe\,{\sc i} signature is in emission, which provides direct evidence of a thermal inversion. These detections suggest that atomic species are common in the atmospheres of ultra-hot Jupiters and have important roles in their energy budgets and climates. We therefore performed a search for the signature of Fe\,{\sc i} in the day-side spectrum of WASP-33b using archival data.

WASP-33b \citep{cameron2010} is one of the hottest ultra-hot Jupiters known with a day-side effective temperature of $>$3100 K \citep[e.g.][]{Smith2011, DeMooij2013, Haynes2015, Zhang2018, vonEssen2020}. It orbits a fast-rotating $\delta$-scuti A5-type star ($v_{\mathrm{rot}\star}\,\mathrm{sin}\,i$= 86\,km\,s$^{-1}$) with a period of $\approx$\,1.22 days. Due to the brightness of the host-star (V\,=\,8.14), WASP-33b is one of the best targets for atmospheric characterization. By analyzing the combined data taken by WFC3/HST, Spitzer and ground-based telescopes, \citet{Haynes2015} detected evidence of stratosphere and a hint of TiO in the atmosphere. This was later confirmed by \citet{Nugroho2017} who directly detected the emission of TiO using high-resolution Doppler-resolved spectroscopy with cross-correlation, which effectively combines thousands of resolved lines, therefore, boosting the signal. This technique makes use of the large Doppler shift from the orbital motion of the planet, enabling us to disentangle the planet's signal from the stationary telluric and (almost stationary) stellar lines. This approach was first demonstrated by \citet{Snellen2010}, and has since become one of the most robust approaches to detect spectral features in the atmospheres of exoplanets \citep[e.g.][]{Hawker2018, Hoeijmakers2018, gibson2020, Turner2020, Stangret2020, Hoeijmakers2020, Pino2020}.

In this letter, we present the detection of Fe\,{\sc i} emission in the day-side of WASP-33b using high-resolution Doppler spectroscopy. In Section \ref{sec:obs&red}, we describe the observations and data reduction. We then describe our modelling of the planetary emission spectrum in Section \ref{sec:modelspec}, and in Section \ref{sec:crosscor}, we detail our search and detection of the Fe\,{\sc i} signal. Finally, in Section \ref{res_and_dis}, we discuss our findings and its implications for the planetary atmosphere.

\section{Observations and Data Reductions} \label{sec:obs&red}
\subsection{Observations and Standard Data Reductions}
Observations of WASP-33b were taken on 26 October 2015 using the High Dispersion Spectrograph \citep[HDS;][]{Noguchi2002} on the Subaru 8.2-m telescope (PID: S15B-090, PI: H. Kawahara). These data were previously presented in \citet{Nugroho2017}, and we refer the reader there for full details. In summary, the data consist of 52 exposures each with an exposure time of 600\,s covering the orbital phase of WASP-33b from $\approx$\,0.206 to 0.538. The observations used a standard NIRc resulting in a wavelength coverage from 6170--8817\,\AA, with a small gap from 7402--7537\,\AA. Image slicer 3 \citep{Tajitsu2012} was used, resulting in an effective slit width of 0.$\arcsec$2, resulting in a spectral resolution of 165,000 (corresponding to $\approx$1.8 km\,s$^{-1}$) sampled at 0.9 km\,s$^{-1}$ per pixel. The extracted spectra are aligned to a common wavelength grid in the telluric frame and are grouped into 2-dimensional arrays with wavelength along one axis and orbital phase along with the other. Each array represents a different spectral order that has been normalized to the continuum profile of the star.

\subsection{Removing Telluric and Stellar Lines}
We removed the telluric and stellar lines using a de-trending algorithm, {\sc SysRem} \citep{Tamuz2005}. {\sc SysRem} was originally developed to fit and remove the systematic effects in the light-curves of large photo-metric surveys and has since been successfully applied to high-resolution Doppler spectroscopy \citep[e.g.][]{Birkby2013, Esteves2017, Nugroho2017,Turner2020, gibson2020, Merrit2020, Nugroho2020}.

We performed the {\sc SysRem} iterations independently to each order directly in flux following the approach in \citet{gibson2020}. The pixel-by-pixel uncertainties were estimated by taking the outer product of the standard deviation of each wavelength bin and exposure which were then normalized by the standard deviation of the data in each order. The best-fit model from each {\sc SysRem} iteration was then summed before dividing out from the data (and corresponding uncertainties). Finally, we applied sigma clipping to identify any strong outliers with a threshold of five times the standard deviation of each wavelength bin, which was then replaced by the mean value of the wavelength bin. Each step of this data processing is shown in Figure 1 for a single order. 

Low-numbers of iterations with {\sc SysRem} will not remove the high-frequency planetary signal (i.e. individual absorption or emission lines) since the radial velocity of the planet changes quickly throughout the observations (from $\approx$\,+213 km s$^{-1}$ to +51 km s$^{-1}$ before the secondary eclipse) while the telluric and stellar lines are relatively stationary in wavelength (although the low-frequency planetary signal, i.e. its continuum, is removed).
However, as the number of {\sc SysRem} iterations increases, the planetary signal will eventually be removed. Moreover, the contamination levels of telluric and stellar lines are different for each spectral order, therefore the optimal number of iterations might vary with the order. Nonetheless, rather than optimize the number of iterations per order (which might artificially enhance the detection significance), we assumed the same number of iterations for all orders. 

\begin{figure}
    \centering
    \includegraphics[width=1.0\linewidth]{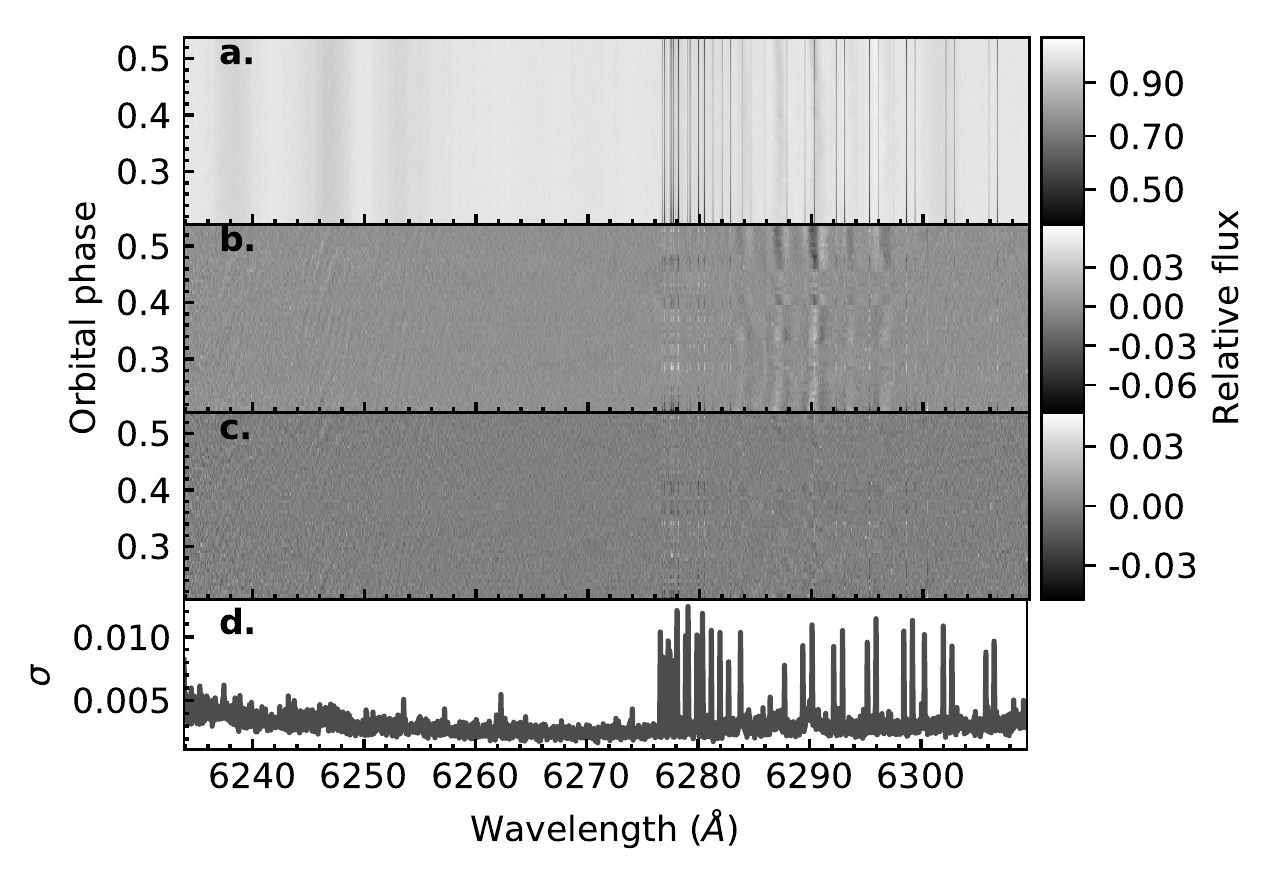}
    \caption{Example of the steps used for telluric and stellar line removal shown for order 3 in the blue. \textbf{(a.)} The reduced and aligned spectra taken from \citet{Nugroho2017}. \textbf{(b.)} The residual spectra after running one iteration of {\sc SysRem}. \textbf{(c.)} Same after six iterations of {\sc SysRem}. \textbf{(d.)} The standard deviation ($\sigma$) of each wavelength bin in the residual spectra.
    \label{fig:sysrem}}
\end{figure} 

\section{Modelling the Planetary Emission Spectrum}
\label{sec:modelspec}

We modeled the planetary spectrum assuming a 1D plane-parallel hydrostatic atmosphere divided into 70 layers evenly spaced in log pressure from 10$^{2}$ to 10$^{-8}$\,bar. The planetary mass and radius were assumed to be 3.266\,M$_{\mathrm{J}}$ and 1.679\,R$_{\mathrm{J}}$, respectively \citep{Kovacs2013}. The temperature-pressure profile of the atmosphere was calculated using equation 29 in \citet{Guillot2010} assuming the intrinsic temperature ($T_{\mathrm{int}}$) of 100 K and $T_{\mathrm{eq}}$ of 3100 K (assuming uniform day-side-only re-radiation). We created a pure inverted atmosphere by setting the visible opacity ($\kappa_{\mathrm{vis}}$) to 2 times the infrared opacity ($\kappa_{\mathrm{ir}}$), while assuming the $\kappa_{\mathrm{ir}}$ of 0.01 cm$^{2}$g$^{-1}$ (e.g. dominated by H$^{-}$ opacity).

We calculated the cross-section of Fe\,{\sc i} for each atmospheric layer using HELIOS-K \citep{Grimm2015} at a resolution of 0.01\,cm$^{-1}$ assuming a Voigt line profile and taking into account natural and thermal broadening only. The line wing cut-off was set to 10$^{8}$ times the Lorentz line width. The line list database and the partition function were taken from \citet{Kurucz2018} and \citet{Barklem2016}, respectively. We also included continuum opacity of bound-free absorption of H$^{-}$ using the equation from \citet{John1988} and Rayleigh scattering by H$_{2}$. The abundances of H$^{-}$ and H$_{2}$, and the mean molecular weight of each atmospheric layers were estimated using {\sc FastChem} \citep{Stock2018}.

The models were calculated in the same way as in \citet{Nugroho2017}, who used them to search for TiO in the day-side of WASP-33b. In total, we produced 21 model templates with Fe\,{\sc i} abundance ranging from a log$_{10}$ volume mixing ratio (VMR) of $-$5.0 to $-$3.0 in steps of 0.1 (assuming constant abundance with altitude), varying the strength of the Fe\,{\sc i} signature relative to the continuum. The resulting spectra were then convolved with a Gaussian kernel\footnote{Using \textsc{pyasl.instrBroadGaussFast}} to the spectral resolution of HDS. To calculate the planet-to-star flux ratio, the spectral models were divided by the flux of the star assuming a black body spectrum with R$_{\star}$ of 1.509\,R$_{\bigodot}$ and $T_{\text{eff}}$ of 7400\,K. Finally, we subtracted the planetary continuum from each model which was estimated using a minimum filter with a window of 4\,\AA. The final result is the line contrast relative to the stellar continuum profile. The planetary spectrum models used in this analysis may be provided upon request.

\section{Searching for the Planetary Signal via Cross-correlation} \label{sec:crosscor}
From our models, the strength of the planetary lines is expected to be at the level of $\approx$\,3--5$\,\times\,$10$^{-4}$ relative to the stellar continuum, therefore the planetary signal is still buried under noise of the residual even after removing the telluric and stellar lines. The Fe\,{\sc i} model has many resolved lines, the signal of which can be combined using cross-correlation to increase the detection significance of the species. To perform the cross-correlation analysis, the template models were Doppler-shifted from $-$200\,km s$^{-1}$ to $+$600\,km\,s$^{-1}$ in steps of 0.2\,km s$^{-1}$, multiplied by the residual spectra after weighting by the variance of each pixel, and finally summed over wavelength for each shift. The cross-correlation function (CCF) is given by
\begin{equation}
    \mathrm{CCF}= \sum \frac{f_{i}m_{i}(v)}{\sigma_{i}^{2}},
\end{equation}
where $f_{i}$ is the mean-subtracted data, $m_{i}$ is the Doppler-shifted mean-subtracted spectrum model to a radial velocity of $v$, $\sigma_{i}$ is the error at $i$ wavelength bin. The sum was first performed over wavelength for each spectral order and stacked into an array. Finally, we summed the CCFs over spectral orders.

To search for the planetary signal, we interpolated the out-of-eclipse CCFs to the planet's rest-frame for a range of possible orbital and systemic velocities, before summing the CCFs over time. Assuming the planet has a circular orbit, the radial velocity of the planet at a given time, RV$_{\mathrm{p}}(t)$, is
\begin{equation}
\mathrm{RV}_{\mathrm{p}} (t)= K_{\mathrm{p}} \sin (2\pi\phi(t)) + v_{\mathrm{sys}} + v_{\mathrm{bary}},
\end{equation}
where $K_{\mathrm{p}}$ is the orbital velocity of the planet, $v_{\mathrm{sys}}$ is the systemic velocity of the star-planet system, $v_{\mathrm{bary}}$ is the barycentric correction, and $\phi(t)$ is the orbital phase of planet at a given $t$ time. We calculated the orbital phase of the planet using the transit epoch taken from \citet{Johnson2015}.

Due to the stellar pulsations (as WASP-33 is a $\delta$-scuti star), the stellar line profiles are distorted as a function of time, and since the stellar spectrum also has Fe\,{\sc i} lines, the pulsation can also be seen in the cross-correlation map between $-v_{\mathrm{rot} \star}\,\mathrm{sin}\,i$+ $v_{\mathrm{sys}}$ and $+v_{\mathrm{rot} \star}\,\mathrm{sin}\,i$+ $v_{\mathrm{sys}}$. Including this pulsation signal in the summation might produce a false-positive detection of the planet's atmosphere, therefore, we only summed up the CCFs from orbital phase of 0.206 to 0.420 for a $K_{\mathrm{p}}$ of $+$200 to $+$400\,km s$^{-1}$ and a $v_{\mathrm{sys}}$ of $-$20 to $+$200\,km s$^{-1}$, both in steps of 0.2\,km\,s$^{-1}$.

The summed CCFs were then stacked into an array with the $K_{\mathrm{p}}$ as the row and $v_{\mathrm{sys}}$ as the column ($K_{\mathrm{p}}-v_{\mathrm{sys}}$ map). To estimate the S/N of the CCFs, we divided the $K_{\mathrm{p}}-v_{\mathrm{sys}}$ map by its noise, estimated from the standard deviation of the CCFs with $K_{\mathrm{p}}>$260\,km s$^{-1}$ and $v_{\mathrm{sys}}>20$\,km\,s$^{-1}$, avoiding the planetary signal at the expected $K_{\mathrm{p}}\approx$ 227--237\,km\,s$^{-1}$ and $v_{\mathrm{sys}}\approx$\,$-3$\,km\,s$^{-1}$ \citep[e.g.][]{cameron2010, Nugroho2017, Yan2019}. We performed this procedure for each {\sc SysRem} iteration (up to 10 iterations) and for each model template.

Finally, we computed the likelihood map following the approach of \citet{gibson2020}, which is a generalised form of the likelihood derived in \citet{Brogi2019}. Here the log likelihood ($\ln \mathcal{L}$) is defined as
\begin{equation}\label{eq:lnlikelihood}
    \ln \mathcal{L}= -\frac{N}{2} \ln \left[\frac{1}{N} \left( \sum\frac{f_{i}^{2}}{\sigma_{i}^{2}} + \alpha^{2}\sum \frac{m_{i}(v)^{2}}{\sigma_{i}^{2}} -2\alpha \mathrm{CCF}\right) \right].
\end{equation}
Here, $N$ is the total number of pixels summed over, and $\alpha$ is the scale factor, which is a multiplicative factor applied to each model, which enables us to recover the strength of the underlying signal relative to the model (i.e. if the maximum likelihood is for $\alpha$ = 1, this implies the input models have the correct overall scaling). We computed the likelihood using $\alpha$ from 0.35 to 1.50 in steps of 0.01. The resulting log-likelihood is a 4-dimensional data-cube with $K_{\mathrm{p}}$, $v_{\mathrm{sys}}$, $\alpha$, and log$_{10}$ VMR as the axes. For numerical reasons, the global maximum value (which coincides with the expected $K_{\mathrm{p}}$ and $v_{\mathrm{sys}}$) was then subtracted (equivalent to normalising the maximum likelihood to 1) and the exponential was computed to produce the 4-dimensional likelihood map. An $\alpha$ value of 0 corresponds to no detection. The significance of the detection was estimated by dividing the median value of the marginalized likelihood of $\alpha$ by its standard deviation.

\section{Results and Discussions} \label{res_and_dis}

We detected Fe\,{\sc i} at S/N\,$\approx$\,6.9 or 6.4-$\sigma$ with $K_{\mathrm{p}}$\,=\,226.0$^{+2.1}_{-2.3}$\,km\,s$^{-1}$ and $v_{\mathrm{sys}}$\,=\,-3.2$^{+2.1}_{-1.8}$\,km\,s$^{-1}$ after 6 passes of {\sc SysRem}, consistent with the expected planetary velocity \citep[e.g. $K_{\mathrm{p}}$\,=\,231.11 $^{+2.20}_{3.97}$\,km\,s$^{-1}$ and $v_{\mathrm{sys}}$\,= -3.02 $\pm$ 0.42\,km\,s$^{-1}$ predicted value using parameters in] [respectively]{Kovacs2013, Nugroho2017}. The detected signal is at its highest after 6 passes of {\sc SysRem} before decreasing, while the constraint parameters remain consistent within 1-$\sigma$ indicating that {\sc SysRem} has a little effect on the planetary signal, therefore we used this number of iterations for the remaining analysis. The marginalized distribution of the likelihood map is shown in Figure 2. The marginalized parameters are summarized in Table \ref{table1}, and the $K_{\mathrm{p}}-v_{\mathrm{sys}}$ map for the best-fit parameters is shown in Figure 3. The smearing effect due to the exposure time used during the observation is expected to be from $\approx$\,0.3\,km\,s$^{-1}$ close to the orbital phase of 0.25, up to $\approx$\,7\,km\,s$^{-1}$ close to the maximum-used orbital phase of 0.42. Assuming the planet is tidally locked, the projected rotational velocity of the planet is $\approx$\,7\,km\,s$^{-1}$. Therefore, the relatively large uncertainties of $K_{\mathrm{p}}$ and $v_{\mathrm{sys}}$ might be due to the combination of both effects. 

\begin{figure}\label{fig:corner}
    \centering
    \includegraphics[width=1.02\linewidth]{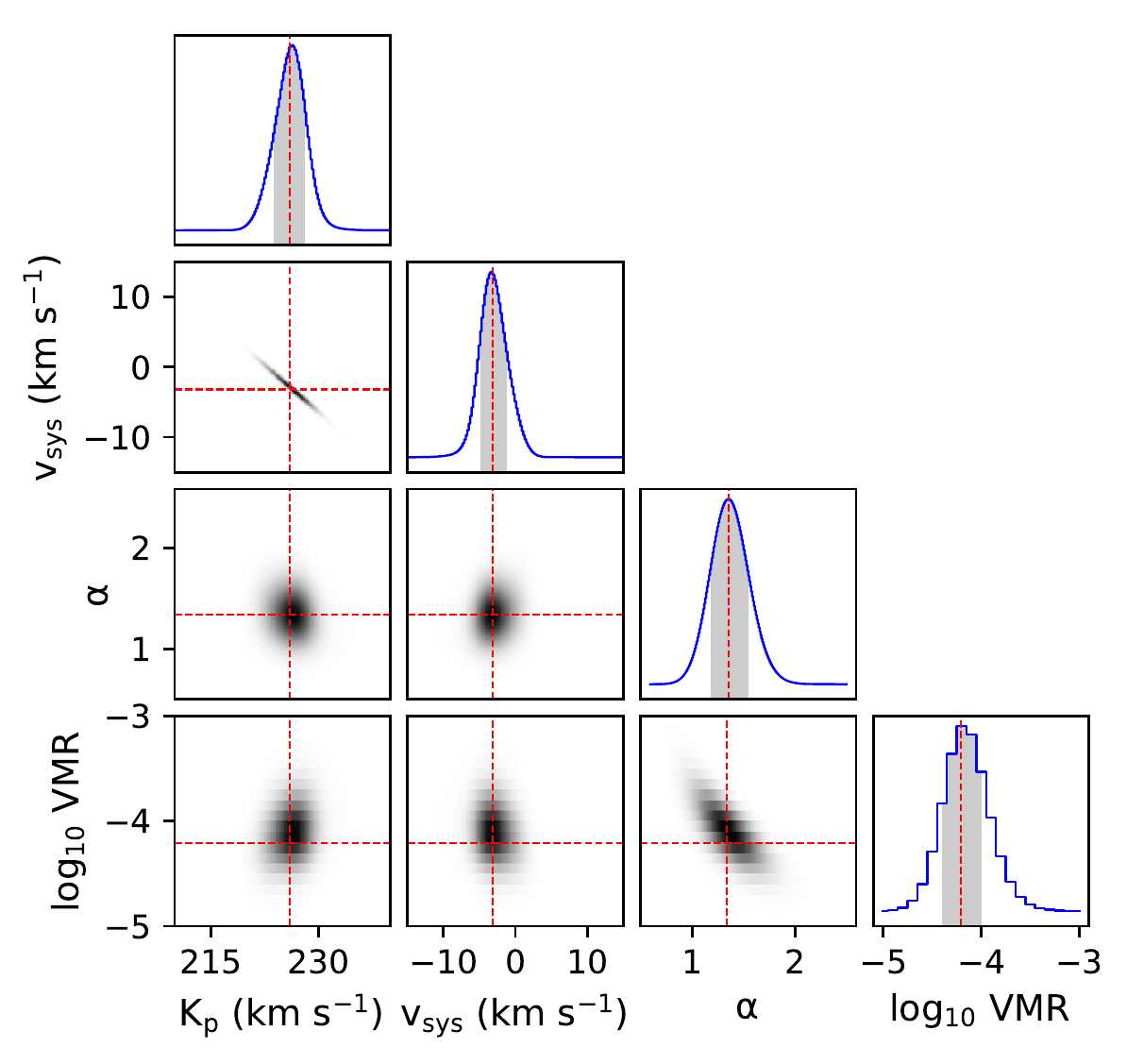}
    \caption{The marginalized likelihood distribution of $K_{\mathrm{p}}$, $v_{\mathrm{sys}}$, $\alpha$, and log$_{10}$ VMR. The red dashed lines indicate the median value of the corresponding distribution. The grey-shaded area indicates the $\pm$1-$\sigma$ limit from the median value.}
\end{figure} 

\begin{table}
\centering 
\caption{The marginalized parameters from the likelihood analysis with $\pm$1-$\sigma$ error. \label{table1}}
\begin{tabular}{lc}
\hline
\hline
Parameter & Value  \\
\hline
$K_{\mathrm{p}}$    (km\,s$^{-1}$) & 226.0$^{+2.1}_{-2.3}$\\
$v_{\mathrm{sys}}$  (km\,s$^{-1}$) & -3.2 $^{+2.1}_{-1.8}$\\
$\alpha$                          & 1.34 $^{+0.21}_{-0.20}$\\
log$_{10}$ VMR                    & -4.2 $\pm$0.2 \\
Significance ($\sigma$)           & $>$6.4\\
\hline\
\end{tabular}
\end{table}

\begin{figure}\label{fig:kpvsys}
\centering
\includegraphics[width=1.02\linewidth]{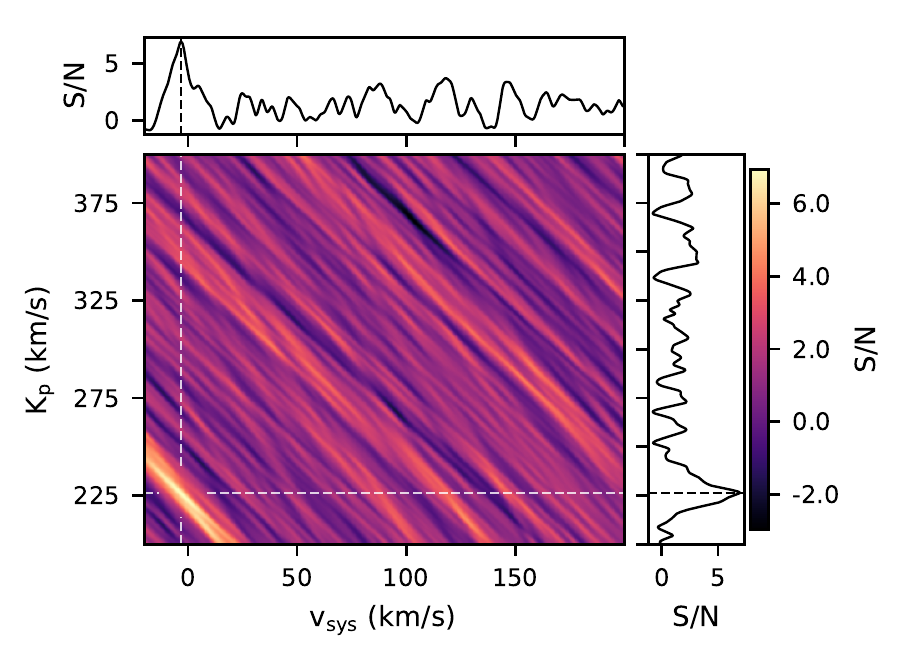}
\caption{The $K_{\mathrm{p}}-v_{\mathrm{sys}}$ map for the best fit parameter calculated by avoiding possible stellar pulsation. The white and black dashed lines show the highest peak in the map with S/N\,$\approx$\,6.9. The black lines in the top and right panels show the CCF along the $K_{\mathrm{p}}$ of 226.0\,km\,s$^{-1}$ and $v_{\mathrm{sys}}$ of $-$3.2\,km\,s$^{-1}$, respectively. The color bar represents the S/N of the map.}

\end{figure}

The corresponding planetary signal can be seen as a dark trail in the CCF map shown by the black arrows before the secondary eclipse (see the left panel of Figure 4a). Only the CCFs inside the solid red lines were used for the summation. As the signals from the stellar pulsation are located outside this area, it is highly unlikely that the detected signal originates from the pulsations. The right panel of Figure 4b shows the mean CCF as a function of orbital phase ($\overline{\mathrm{CCF}}$). We can see that before the orbital phase of 0.25, the detected planetary signal is relatively weak. This is expected for a tidally locked planet where the heat distribution from the day-side to the night side is inefficient, raising the temperature contrast between both hemispheres as it was measured using Spitzer and TESS phase-curve by \citet{Zhang2018} and \citet{vonEssen2020}, respectively.

Using our likelihood analysis, we also found that the best fit spectrum has the Fe\,{\sc i} abundance of log$_{10}$ VMR= -4.2$\pm$0.2, however, since the abundance of chemical species degenerates with the atmospheric temperature profile \citep[e.g.][]{Madhu2009}, our retrieved abundance of Fe\,{\sc i} is only correct for the specific atmospheric profile that we assumed and should be interpreted cautiously. Nonetheless, it is still very helpful to investigate whether the atmosphere has thermal inversion or not as high-resolution spectroscopy analysis is highly sensitive to the line profile which is model-independent \citep[e.g.][]{Schwarz2015}. Figure 4 shows the best-fit model and the corresponding contribution function. In our model, for an atmosphere with log$_{10}$ VMR of Fe\,{\sc i}= -4.2, the planetary signal is mostly coming from the atmospheric layers within the pressure of 1-0.001 bar with the continuum profile mostly from H$^{-}$ opacity. Around this pressure, the temperature profile of the planet is inverted, therefore all of the Fe\,{\sc i} lines in the spectrum model are in emission. Therefore, our result also confirms the existence of a thermal inversion layer in the day-side of WASP-33b that has been detected previously by \citet{Haynes2015} and \citet{Nugroho2017}.

The scaled line contrast of the best fit spectrum model seems to be consistent with the secondary eclipse depth measured by \citet{vonEssen2020} from the analysis of TESS phase-curve. The retrieved $\alpha$ value showed that our model has slightly under-estimated the strength of the line contrast of the planet. This can be interpreted that either the atmosphere that we probed has a higher temperature or it has stronger inversion than our model (higher lapse rate) or we have overestimated the H$^{-}$ abundance. Breaking this degeneracy can be done by varying the assumed temperature profile, however, this would be computationally expensive. As the main purpose is to detect Fe\,{\sc i}, fixing the temperature profile and the continuum opacity (which is assumed to be dominated by the bound-free H$^{-}$) and varying the VMR allows us to explore different temperature regime to some extent. Changes in the VMR result in the atmosphere becoming optically thick at different pressures, and therefore we probe different temperatures in our assumed atmospheric model while minimizing free parameters. More complex and detailed analysis is beyond the scope of this letter.

This detection tends to suggest that Fe\,{\sc i} is common in the atmosphere of hot Jupiters and could be the main caused for temperature inversion as it was suggested observationally for WASP-121b \citep{gibson2020, Merrit2020} and theoretically by \citet{Lothringer2018, Lothringer2019}. However, for WASP-33b, the presence of both Fe\,{\sc i} and TiO in its atmosphere might indicate that both species contribute to creating the observed stratosphere. This provides a unique opportunity to directly study their relative contributions to the overall optical opacity, e.g. by constraining their relative abundances, and, therefore, to the energy budget of the atmosphere. However, this requires a much more detailed exploration of different atmospheric profiles, and any further interpretation will be the subject for future study.

\begin{figure*}\label{fig:ccmapnspec}
\gridline{\fig{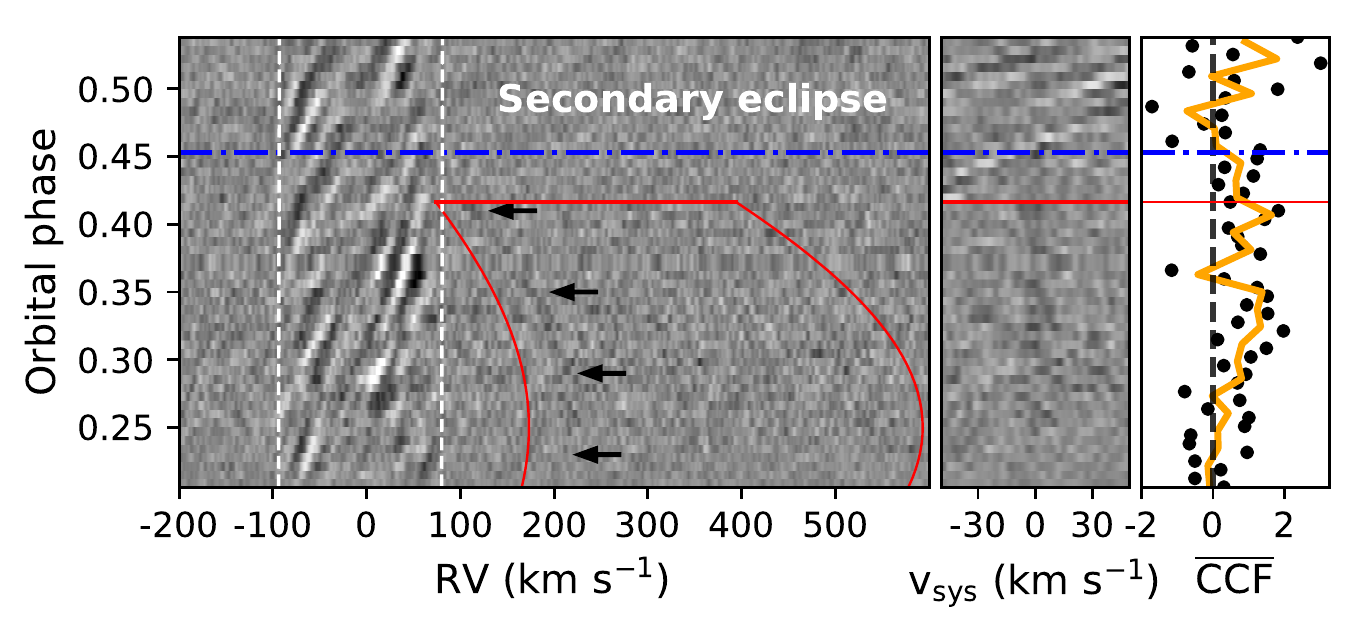}{0.55\linewidth}{(a)}
\fig{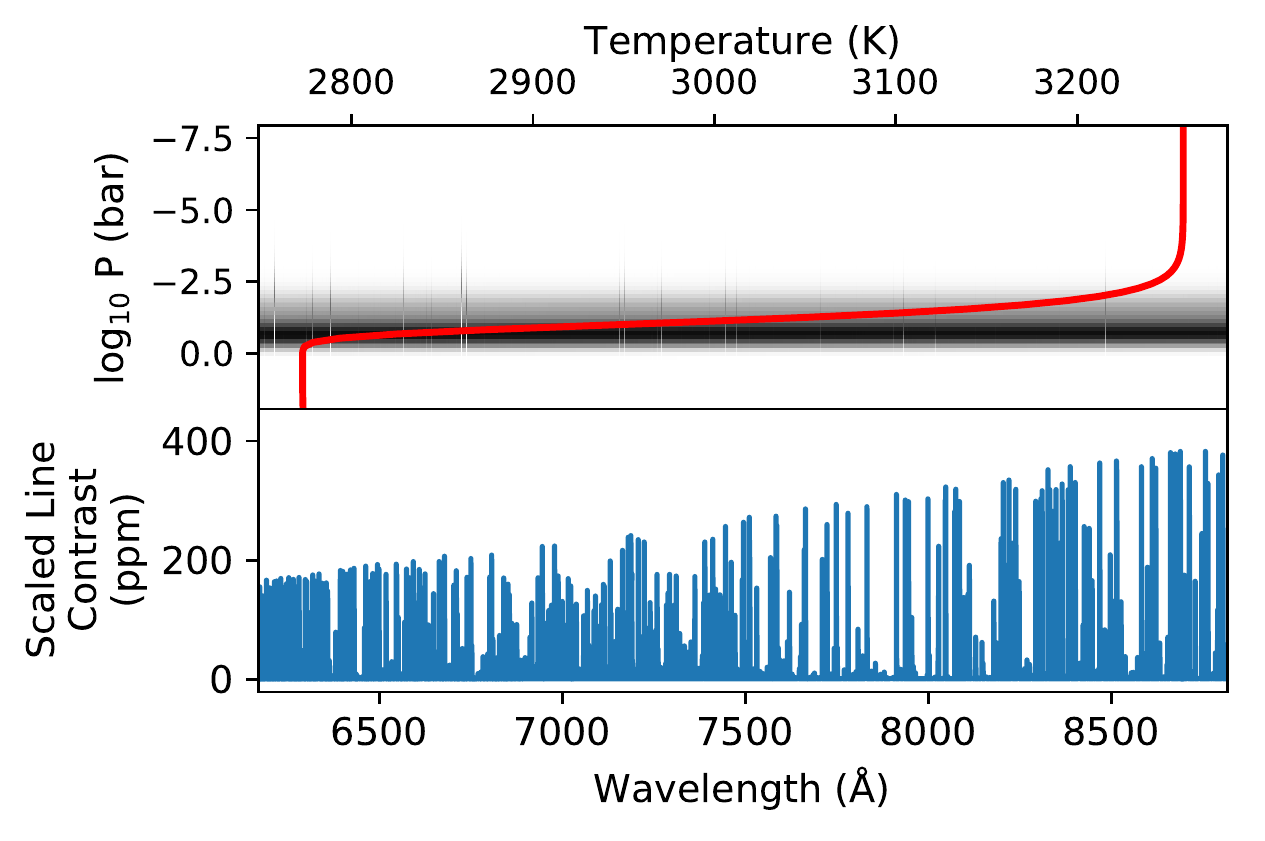}{0.45\linewidth}{(b)}}
\caption{\textbf{(a)} \textit{left panel}: The CCF map of Fe\,{\sc i} with log$_{10}$ VMR of -4.2 after 6 {\sc SysRem} iterations. The white dashed line marks the projected rotational velocity of the star. The blue dot-dashed line marks the beginning of the secondary eclipse. Four black arrows point to the location of the detected signal. Only the area inside the solid red lines was used when summing over the CCF. \textit{middle panel}: The CCFs at the planetary rest-frame. \textit{right panel}: The black dots show the mean CCF calculated for $\pm$4\,km s$^{-1}$ from the center of the planet's signal. The orange line shows the binned-CCF by 2 exposures. \textbf{(b)} The best fit spectrum of Fe\,{\sc i} with log$_{10}$ VMR of -4.2. The top panel shows the contribution function in the wavelength range of HDS represented by the gray shades. The red line represents the temperature profile that we adopted in our modeling. The bottom panel shows the best-fit model of Fe\,{\sc i}.}
\end{figure*}

\section*{Acknowledgements}
We are extremely grateful to the anonymous referee for constructive and insightful comments that greatly improved the quality of this letter. This work is based on data collected at Subaru Telescope, which is operated by the National Astronomical Observatory of Japan. S.K.N. and C.A.W. would like to acknowledge support from UK Science Technology and Facility Council grant ST/P000312/1. N. P. G. gratefully acknowledges support from Science Foundation Ireland and the Royal Society in the form of a University Research Fellowship. M.K.H. is supported by funding from the Natural Sciences and Engineering Research Council (NSERC) of Canada. H.K. is supported by a Grant-in-Aid from JSPS (Japan Society for the Promotion of Science), Nos. JP18H04577, JP18H01247, and JP20H00170. This work was also supported by the JSPS Core-to-Core Program "Planet$^{2}$" and SATELLITE Research from Astrobiology Center (AB022006). We are also grateful to the developers of the {\sc Numpy}, {\sc Scipy}, {\sc Matplotlib}, {\sc Jupyter Notebook}, and {\sc Astropy} packages, which were used extensively in this work \citep{2020SciPy-NMeth, Hunter:2007, Kluyver:2016aa, astropy:2013, astropy:2018}.


\bibliography{FeIW33b-nugroho}{}
\bibliographystyle{aasjournal}



\end{document}